\documentstyle[preprint,amsfonts,amsmath,prd,aps,12pt,epsf]{revtex}
\def\vvrh{\boldsymbol\varrho}
\def\vvel{\boldsymbol\varepsilon}
\def\vell{\boldsymbol\ell}
\begin{document}
\preprint{
\vbox{
\halign{&##\hfil\cr
         & DESY 98-164\cr
         & hep-ph/9910531\cr
         & October 1999 \cr
         & \cr
         & \cr
}}}
\vskip 0.5cm
\title{The Photon Wave Function in Non-forward Diffractive Scattering with
Non-vanishing Quark Masses}
\author{S. Gieseke\footnote{Supported by Graduiertenkolleg `Theoretische
Elementarteilchenphysik' and TMR-Net\-work `QCD and Particle Structure',
contract number FMRX-CT98-0194 (DG 12 - MIHT).} 
and Cong-Feng Qiao\footnote{Alexander von Humboldt Fellow.}}
\vskip 6pt
\address{II Institut f\"ur  Theoretische Physik, Universit\"at
Hamburg,\\Luruper Chaussee 149, D-22761 Hamburg, Germany.}
\maketitle
\vskip 9mm
\begin{abstract}
The light-cone Photon wave function in explicit helicity states, valid for
massive quarks and in both momentum and configuration space, is presented 
by considering the leading order photon-proton hard scattering, i.e., the 
splitting quark pair scatters with the proton in the Regge limit. Further
we apply it to the diffractive scattering at nonzero momentum transfer and
reach a similar factorization as in the case of zero momentum transfer.

\vspace{4mm}
\noindent 
PACS number(s): 12.38.Bx, 12.40.Nn, 13.60.-r
\end{abstract}
~\\
\vfill \eject
\section{Introduction}

The hard diffractive photon-proton scattering are of particular interest
in testing the perturbative QCD(pQCD) because of not only the HERA
prolific experimental data but also the theoretical developments to date.  
Diffractive processes generally characterized by the presence of large
rapidity gaps in the hadronic final state are conventionally ascribed to
Pomeron exchange. Before the advent of QCD, hard diffractive processes are
well described by Regge theory \cite{coll}. Although the relation between
Regge asymptotics with pQCD description, the well-known BFKL equation
\cite{bfkl}, is still unclear, we have seen the possibility of gaining an
understanding of the nature of diffraction, the Pomeron and its coupling
to partons from the first principle. The hard diffractive processes in
deep-inelastic scattering(DIS) are of particular interest toward these
goals, in which people believe could reach the transition region between
perturbative and non-perturbative strong interactions.

It is believed that in DIS the fast-moving photon dissociates into partons
long before its interacting with the target and the dissociation of the
photon into partons is described by the so-called photon wave function
\cite{nz,ahm,bfgms,bekw}. In the beginning of the scattering, the photon
splits into a quark-antiquark pair, which may or may not evolve into more
partons by strong interaction before it reaches the target proton
\cite{bekw}. The simplest process in diffractive DIS is of the $q\bar{q}$
production, which is conveniently described in terms of light-cone photon
wave functions procedure in the proton rest frame. Because in DIS $Q^2$,
the photon virtuality, is relatively large, the quark-antiquark pair in
impact parameter space can be referred as a color dipole. The diffractive
interaction of the color dipole with the proton is mediated by Pomeron,
which may be described by, e.g., two gluon model \cite{ln}.
 
The photon wave function has been introduced and used for many years(for a
recent paper related to it, see e.g. \cite{hks}), however we find that in
amplitude level the photon-quark-antiquark wave function with
non-vanishing quark masses are not explicitly shown in the helicity basis.
In \cite{ahm} expressions for massless quarks and transverse photons are
given explicitly in terms of helicity amplitudes. \cite{nz,NZC49} give
expressions for the squared wave functions, including longitudinal photons
and also massive quarks. A derivation of the amplitudes in terms of an
explicit spinor representation limited to massless quarks is given in
\cite{IvaWuest}.  For completeness, in this paper we will reconsider a
perturbative derivation of the photon wave function, in explicit helicity
basis and including fermions masses, by considering the amplitude for the
diffractive process $\gamma^*p\to p+X$ both in momentum and configuration
space. In practice, we consider the kinematic limit, where the $\gamma^*p$
energy $s$ is much larger than the diffractive mass $M_X^2$, the relative
transverse momentum squared of the $q\bar q$-pair $\bf{k}^2$, the
virtuality of the photon $Q^2$ and the momentum transfer $t$ from the
virtual photon system to the proton. In this limit the amplitude is
factorized basically into the photon wave function and the unintegrated
off-diagonal gluon structure function as expected.

Instead of the space of transverse momenta one can transform this
amplitude to configuration space, where the conjugate variable to the
transverse momentum separation of the $q\bar q$ pair is its transverse
size. This leads to the convenient dipole-picture in the proton rest
frame, saying that the virtual photon fluctuates into the $q\bar q$-pair
long before it undergoes the interaction with the proton and therefore
this fluctuation and the interaction of the dipole with the proton are
independent and hence factorize in the transverse space.  In this way a
description of cross sections for inclusive diffractive scattering
processes, as well as of diffractive vector meson production, can be
carried out completely in terms of the photon wave function and the so
called dipole cross section, see e.g. \cite{wuest,Golec,FKS}.

The paper is organized as follows. In section II we will derive the photon
wave function in momentum space by using the helicity method developed in
\cite{bm}. Following we will transform the wave function into
configuration space and to see the factorized form of photon-proton
interaction in two-gluon model. In the end there will be a brief summary.

\section{Light-Cone Wave Functions in Momentum Space} 
 
The photon wave function is a very useful tool in the calculation of a
high energy virtual photon scattering off the target. One of the nice
features of it is that it can not only being used for single gluon
exchange processes (e.g, in determining $F_2$), but can be extended to
include multi-gluon exchange processes. Here, before considering the
amplitudes for diffractive scattering, we calculate the photon wave
function from a simpler process, where a single gluon is exchanged in the
$t$-channel as depicted in Fig.1.

Throughout the calculation we will use the Sudakov decomposition for all
momenta with respect to the light cone vectors $q' = q - (Q^2/s) p_B$
and $p'= p_B + (m_B^2/s) q'$($s = 2 p'\cdot q'$; $q$ is the momentum
of the photon with virtuality $Q^2=-q^2$ and $p_B$ is the momentum of
the incoming quark $B$ with mass $m_B$). For example, in diagram (a) the 
momenta $k$ of internal quark line and $r$ of the exchanged gluon 
have the following Sudakov decomposition:
\begin{eqnarray}
  \label{eq:sudakov}
  k &= \alpha q' + \beta p' + k_\perp,\\
  r &= \frac{t}{s} q' + x_{\rm{I\!P}} p' + r_\perp.
\end{eqnarray}
In this way the antiquark and quark carry fractions $\alpha$ and
$(1-\alpha)$ in the $q'$-direction. Noted that in the following we 
often use the Euclidean form of the transverse momenta marked in boldface, 
i.e.\ $k_\perp^2 = -\bf{k}^2$. 

>From the on-shell conditions for the outgoing quarks, together with the 
assumptions that $Q^2\sim |r^2=t|\sim M_X^2 \ll s$, we know
$x_{\rm{I\!P}}$ and $\beta$ to be of the order $t/s$. In this limit we use
the usual decomposition of the metric tensor 
\begin{equation}
  \label{eq:metrtens}
  g_{\mu\nu} = \frac{2}{s}(p'_\mu q'_\nu + p'_\nu q'_\mu) +
  g_{\mu\nu}^\perp 
\end{equation}
for the gluon propagator and retain only the first term with an
appropriate contraction, while the other terms are suppressed by
powers of $t/s$. Furthermore, the denominators of the quark lines in
diagrams (a) and (b) can be expressed respectively as
\begin{eqnarray}
  \label{eq:denoms}
&&\Delta_a = k^2-m^2 =-\frac{1}{1-\alpha}[\alpha(1-\alpha)Q^2 + {\bf k}^2
+ m^2] \equiv -\frac{1}{1-\alpha} D({\bf k}), \\ 
&&\Delta_b = (q-k-r)^2-m^2 = -\frac{1}{\alpha}[\alpha(1-\alpha)Q^2 +
({\bf k} + {\bf r})^2 + m^2]\equiv -\frac{1}{\alpha} D({\bf k}+{\bf r}).
\end{eqnarray}
The mass in the upper fermion line is always denoted as $m$. 

After these denotations it is quite straightforward to write down the 
amplitude, by virtual of the helicity method, e.g. in \cite{bm},  
for one gluon exchange as 
\begin{equation}
  \label{eq:ampl}
  {\cal A} = - ee_f{\cal C} \frac{2g^2 \delta_{\lambda_B, \lambda_{B'}}}{t} 
    \bar u_{\lambda'}(p_{A'}) 
    \left(\frac{\chi_a}{\Delta_a} + \frac{\chi_b}{\Delta_b} \right) v_\lambda
    (p_A) 
\end{equation}
with 
\begin{equation}
  \label{eq:chis}
  \chi_a = \!\not p' (\!\not k + m) \!\not \varepsilon^\gamma,
  \qquad \chi_b = - \!\not \varepsilon^\gamma  (\!\not q - \!\not k - \!\not r
  -  m) \!\not p', 
\end{equation}
and $\lambda' (\lambda) = \pm$ denotes the helicity of the (anti-)quark.  
${\cal C}$ is the color factor and $g$ is the strong coupling constant.
The photon polarization vectors in Eq.(\ref{eq:chis}) can be chosen as
\begin{equation}
  \label{eq:polavec}
  \varepsilon^0 = \frac{1}{Q}(q' + x_B p'), \qquad
  \varepsilon^(\pm)_\perp = \frac{1}{\sqrt{2}}(0,1,\gamma i, 0), 
\end{equation}
denoting the longitudinal($\gamma = 0$) and transverse($\gamma=\pm$) cases, 
respectively.

Still using the helicity method and keeping the mass term one reaches
the helicity expressions for longitudinal and transverse polarizations 
of the photon in diagram (a)(and a similar one for diagram (b)),   
\begin{eqnarray}
  \label{eq:chiafter}
  \chi_a (\gamma=0) &= \alpha s \!\not \varepsilon^0 
  + \frac{s}{2Q}\!\not r_\perp
  - i\epsilon(\mu,\varepsilon^0,p_{B'},p_B)\gamma_\mu\gamma_5, \\
  \chi_a (\gamma=\pm) &= \alpha s \!\not \varepsilon_\perp 
  + {\boldsymbol \varepsilon}_\perp\cdot{\bf r}\!\not p'
  - i\epsilon(\mu,\varepsilon_\perp,p_{B'},p_B)\gamma_\mu\gamma_5. 
\end{eqnarray}
Here only leading terms in the high energy limit are kept. With
(\ref{eq:chiafter}) we obtain the helicity amplitude for (\ref{eq:ampl}), 
which can be written in a compact form
\begin{equation}
  \label{eq:full}
  {\cal A} = eg^2\frac{2s}{t}\delta_{\lambda_B, \lambda_{B'}} {\cal
  C} \sqrt{\alpha(1-\alpha)}\Big(\Psi({\bf k}, \alpha)-\Psi({\bf k}+{\bf r},
\alpha) \Big). 
\end{equation}
Here, $\Psi({\bf k},\alpha)$ is the so called light-cone photon wave function,
which contains all dependence of the amplitude on the properties of the
virtual photon and the $q\bar q$-pair in helicity basis.
\begin{eqnarray}
  \label{eq:wflong}
  \Psi^0_{\pm\mp}({\bf k}, \alpha) &= \frac{2e_f\alpha(1-\alpha)Q}
  {\alpha (1-\alpha) Q^2 + {\bf k}^2 + m^2},\\ 
  \label{eq:wftrans1}
\Psi^\pm_{\pm\mp}({\bf k}, \alpha) &= \frac{\sqrt{2}e_f\alpha \underline k}
{\alpha (1-\alpha) Q^2 + {\bf k}^2 + m^2},\\ 
\label{eq:wftrans2}
\Psi^\pm_{\mp\pm}({\bf k}, \alpha) &= \frac{-\sqrt{2}e_f(1-\alpha) \underline k}
  {\alpha (1-\alpha) Q^2 + {\bf k}^2 + m^2},\\ 
  \label{eq:wftransfl}
  \Psi^\pm_{\pm\pm}({\bf k}, \alpha) &= \frac{\sqrt{2}e_fim}
  {\alpha (1-\alpha) Q^2 + {\bf k}^2 + m^2},\\
  \label{eq:wftransm}
  \Psi^0_{\pm\pm}({\bf k}, \alpha) &= \Psi^\pm_{\mp\mp}({\bf k},\alpha) = 0,
\end{eqnarray}
where $\underline k = k_x + i\gamma k_y$. Note that in the additional term 
(\ref{eq:wftransfl}), appearing only for massive quarks, there is a naive
helicity conservation $\gamma=(\lambda+\lambda')/2$. A similar situation
was found for (\ref{eq:wflong}) but there the mass dependence always
cancels in the difference of the wave functions. On the other hand, in the
massless limit we can only have the configuration $\lambda = -\lambda'$
within the $q\bar q$-pair.  This behavior is easily understood from the
appropriate limit in $m/\sqrt{s}$: in the high energy limit the fermions'
helicities are always opposite and independent of the coupling, in
contrast to the nonrelativistic limit where we have helicity conservation.
Besides, in massless limit we find a agreement of the expressions
(\ref{eq:full})-(\ref{eq:wftransm}) with references \cite{ahm,wuest}.

In the following we apply the obtained photon wave functions to the 
diffractive scattering in two gluon exchange model as depicted in Fig.2.
In the leading $\log x$ approximation the amplitude is dominated by
its imaginary part, from which we will reconstruct the full amplitude
later on. By virtue of the Cutkosky rules we may consider each diagram as
a composition of two, where the crossed fermion lines shown in Fig.2 are
on mass shell. The left part has a structure as in the one gluon exchange
discussed above and the right part is just a simple one gluon exchange
between two quark lines. From the cuts in the fermion lines we have the
integration loop momentum
\begin{equation}
  \label{eq:loop}
  \ell = \alpha_\ell q' + \beta_\ell p' + \ell_\perp,
\end{equation}
where from the mass-shell conditions we know $\alpha_l$ and $\beta_l$ 
are both of order $t/s$.

>From the respective right parts of the diagrams we get a factor
$2\alpha s/({\bf r} - {\vell})^2$ or 
$2(1-\alpha) s/({\bf r}-{\vell})^2$ 
with helicity conservation within the fermion
lines, depending on whether the right gluon couples to the quark or
the antiquark. On the other hand we get factors $1/2\alpha s$ or
$1/2(1-\alpha) s$ from the integration over the longitudinal part of
the loop momentum and the on-shell conditions, therefore we are only
left with the denominators from the right parts of the diagrams. 

To determine the left parts, we can simply read off the transverse
momentum of the virtual quark line from the diagrams and get a photon
wave function with this argument, similar to the one gluon case, 
but now we also have the transverse part of the loop momentum as an
argument of the wave functions. Furthermore we now have a different
color structure. Putting everything together, we can  write the full 
amplitude, arising from the diagrams in Fig.2, as a 'double difference'
\begin{eqnarray}
  \label{eq:diffmom}
  {\cal A}&=&i eg^4\frac{s}{t}\delta_{\lambda_B, \lambda_{B'}} 
  {\cal C}' \sqrt{\alpha(1-\alpha)} \int \frac{d^2{\vell}}{(2\pi)^2}
  \frac{{\bf r}^2}{{\vell}^2({\bf r- l})^2} \\ 
  &\times& \Big\{\Psi({\bf k}, \alpha) + \Psi({\bf k} + {\bf r}, \alpha) 
  - \Psi({\bf k} + {\vell}, \alpha)
  - \Psi({\bf k}+ {\bf r} - {\vell}, \alpha) \Big\},
  \nonumber
\end{eqnarray}
with the wave function $\Psi({\bf k}, \alpha)$ defined as in Eqns.
(\ref{eq:wflong}) -- (\ref{eq:wftransm}). This kind of 'double difference'
was already obtained in \cite{bfgms} and \cite{bekw} in a proper limit.

To include the non-perturbative coupling of the two gluons to the
proton, one needs to go beyond the leading order and meets the
unintegrated off-diagonal gluon distribution
${\cal{F}}(x,x',{\vell}^2,{\bf r}^2)$, a suitable generalisation of the
unintegrated gluon distribution to the off-diagonal case, for more see
e.g. \cite{GKM} and references therein for discussions of off-diagonal
gluon distributions in diffraction. Here, $x$ and $x'$ denote the
longitudinal momentum fractions of the two gluons, coupling to the proton
and $x_{\rm{I\!P}}=x-x'$. Integrating over ${\vell}$ would give an
off-diagonal gluon distribution,
\begin{equation}
  \label{eq:strudef}
  \int^{Q^2}\!\!d^2{\vell} {\cal{F}}(x,x',{\vell}^2,{\bf r}^2)
  = G(x, x', {\bf r}^2, Q^2)    
\end{equation}
and in the limit $x\approx x'$ we have $G(x,x,{\bf r}^2=0,Q^2)=xg(x,Q^2)$, 
with $g(x,Q^2)$ being a diagonal gluon distribution. In \cite{MR,FFGS} the
deviation of $G(x,x',t,Q^2)$ from $xg(x,Q^2)$ is discussed in detail. 

Now we may write the general amplitude for diffractive scattering off
the proton as
\begin{equation}
  \label{eq:fulldiff}
{\cal A} =\,i \frac{\pi}{4}eg^2 \sqrt{\alpha(1-\alpha)} s \int
\frac{d^2{\vell}}{\pi {\vell}^2} {\cal{F}}(x,x',{\vell}^2,{\bf r}^2) 
D\Psi({\bf k}, {\bf r},{\vell}),  
\end{equation}
where the shorthand notation 
\begin{equation}
D\Psi({\bf k}, {\bf r}, {\vell}) = \Psi({\bf k}, \alpha)+\Psi({\bf k} 
+{\bf r}, \alpha)-\Psi({\bf k}+{\vell}, \alpha) - \Psi({\bf k}+{\bf r}
- {\vell}, \alpha)
\end{equation}
for the double difference of the wave functions, 
${\cal{F}}(x,x',{\vell}^2,{\bf r}^2)$ is normalized as in the diagonal 
case, i.e. 
${\cal{F}}(x,x,{\vell}^2,{\bf r}^2 =0) = {\cal{F}}(x,{\vell}^2)$, 
with ${\cal{F}}(x,{\vell}^2)$ being the unintegrated gluon structure 
function \cite{bekw}. Furthermore, we left the dependence on $t$ in the 
unintegrated  structure function to keep the express more general.

\section{Expressions in Configuration Space}

It is more clear to consider the factorization of diffractive amplitudes
in configuration space. Therefore we will proceed the Fourier transform in
this section with respect to the transverse momenta ${\bf k}$ and ${\bf
r}$. The variable conjugate to ${\bf k}$ is the transverse separation of
the $q\bar q$-pair ${\vvrh}$, or simply the called `dipole size'.
Similarly, the variable conjugated to the momentum transfer between the
diffractive system and the proton is the impact parameter ${\bf b}$.

The conjugated photon wave functions are
$\psi^\gamma_{\lambda'\lambda}({\vvrh})$ in ${\vvrh}$-space as 
\begin{eqnarray}
  \label{eq:cswflo}
\psi^0_{\pm\mp}({\vvrh}) &=&\frac{1}{\pi}e_f\alpha(1-\alpha) Q
K_0(\delta\varrho),\\
  \label{eq:cswftr1}
\psi^\pm_{\pm\mp}({\vvrh}) &=&\frac{i}{\pi}e_f\alpha\delta 
\frac{{\vvrh}\cdot{\vvel}}{\varrho}K_1(\delta\varrho),\\
  \label{eq:cswftr2}
\psi^\pm_{\mp\pm}({\vvrh}) &=& -\frac{i}{\pi}e_f(1-\alpha)\delta 
\frac{{\vvrh}\cdot{\vvel}}{\varrho}K_1(\delta\varrho),\\
  \label{eq:cswftrfl}
\psi^\pm_{\pm\pm}({\vvrh}) &=&\frac{im}{2\pi}e_f K_0(\delta\varrho),
\end{eqnarray}
where we used $\delta^2=\alpha(1-\alpha)Q^2+m^2$ as a shorthand notation
and $K_\nu(z)$ is modified Bessel function. After squaring, our results
agree with that of \cite{NZC49}. Using these wave functions, we can easily
transform the diffractive amplitude (\ref{eq:fulldiff}) into configuration
space, 
\begin{eqnarray}
  \label{eq:ftampl}
\tilde{\cal A}^D({\vvrh},{\bf b})
&=&\int\!\frac{d^2{\bf k}}{(2\pi)^2}\int\!\frac{d^2{\bf r}}{(2\pi)^2}
e^{i{\bf k}\cdot{\vvrh}}e^{i{\bf r} \cdot {\bf b}}
{\cal A}^D({\bf k},{\bf r})\nonumber\\
&=&{\cal B} \int \!\frac{d^2{\bf k}}{(2\pi)^2}  \int\!\frac{d^2{\bf r}}{(2\pi)^2}
e^{i{\bf k} \cdot {\vvrh}}  e^{i{\bf r} \cdot {\bf b}}\int\!
\frac{d^2{\vell}}{\pi{\vell}^2}\alpha_s(\mu^2){\cal{F}}(x,x',{\vell}^2,
{\bf r}^2) D\Psi({\bf k},{\bf r}, {\vell}),  
\end{eqnarray}
where ${\cal B}$ contains the factors in front of the integral in
(\ref{eq:fulldiff}).  In doing the ${\bf k}$ integration we make use of the
wave function (\ref{eq:cswflo})-(\ref{eq:cswftrfl}) and pick up 
appropriate phases from a shift in the integration region, get 
\begin{equation}
  \label{eq:wftrafo}
  \int \!\frac{d^2{\bf k}}{(2\pi)^2}e^{i{\bf k} \cdot {\vvrh}}
  D\Psi({\bf k}, {\bf r}, {\vell}) = \psi({\vvrh})
  [1+e^{-i{\bf r}\cdot {\vvrh}}
  -e^{-i{\vell}\cdot {\vvrh}}
  -e^{-i({\bf r}-{\vell}) \cdot{\vvrh}}].
\end{equation}
Therefore, we may write 
\begin{eqnarray}
\tilde{\cal A}^D({\vvrh},{\bf b}) &=& B\psi({\vvrh}, \alpha)
\int \!\frac{d^2{\bf r}}{(2\pi)^2} e^{i{\bf r} \cdot {\bf b}}\int\!
\frac{d^2{\vell}}{\pi{\vell}^2}\nonumber\\
&\times&\alpha_s(\mu^2){\cal{F}}(x,x',{\vell}^2,{\bf r}^2) 
[1-e^{-i{\vell}\cdot {\vvrh}}]
[1-e^{-i({\bf r}-{\vell})\cdot{\vvrh}}]
\nonumber\\
&\equiv&\psi({\vvrh},\alpha)\int \!\frac{d^2{\bf r}}{(2\pi)^2}
e^{i{\bf r} \cdot {\bf b}}\sigma_{q\bar q}({\vvrh},{\bf r}),
\end{eqnarray}
where we have put the ${\vell}$-dependent part and the factor into the 
definition of the proton-dipole cross section. Carring out the last 
Fourier transform from momentum transfer to impact parameter, we get 
the expected factorized amplitude, as in the zero momentum case in e.g.
\cite{bjw}, 
\begin{eqnarray}
  \tilde{\cal A}^D({\vvrh}, {\bf b}) &=
  \psi({\vvrh}, \alpha)
  \sigma_{q\bar q}({\vvrh},{\bf b})
\end{eqnarray}
in the case of nonzero momentum transfer. We see that the the physical 
picture, where the fluctuation of the virtual photon into the $q\bar q$ pair 
occurs long before the interaction of dipole and proton, still holds.  
But for now, the dipole proton cross section 
$\sigma_{q\bar q}({\vvrh},{\bf b})$ depends on not only the dipole
size ${\vvrh}$, but also the impact parameter ${\bf b}$.

\section{Summary}

In the diffractive scattering of a virtual photon off a target it is
technically useful to introduce the photon wave function in the
calculation.  In this paper we have recalculated the wave functions
including massive quarks in the explicit helicity basis, which is not
fully and explicitly presented in the literature by now.

On the other hand, the concept of the photon wave function turned out to
arise naturally in considering the diffractive scattering, where the
factorized part of the photon is expressed in terms of the wave function.  
We get know that in addition to the dipole size ${\vvrh}$, the
dipole cross section also depends on the impact parameter ${\bf b}$ of
virtual photon and proton in transverse configuration space. It is shown
manifestedly that the factorization of the diffractive amplitude into the
photon wave function and the unintegrated structure function still holds
in the case of non-zero momentum transfer, which is widespreadly believed.

\vskip 1.2cm
\centerline{\bf ACKNOWLEDEMENTS}
\vskip 0.3cm
We are grateful to J.~Bartels for many helpful discussions and 
suggestions.

\end{document}